\documentclass[aps,prd,a4paper,nofootinbib,
preprintnumbers,superscriptaddress,showpacs,showkeys,
twocolumn]{revtex4-2}

\usepackage{graphicx} 
\usepackage{bm} 
\usepackage{amsmath}
\usepackage{slashed}

\usepackage{xcolor}
\usepackage[
colorlinks=true,
linkcolor=blue,
citecolor=blue,
urlcolor=blue,
breaklinks,
pdfstartview=FitH,
bookmarksopenlevel=0,
bookmarks=true,
pdftoolbar=true,
pdfauthor={{}}
pdftitle={Reanalysis of critical exponents for the O(N) model via a hydrodynamic approach to the Functional Renormalization Group},
pdfkeywords={FRG, fluid dynamics, critical exponents, effective field theories, QCD}
]{hyperref}

\usepackage[sort,compress]{cleveref}
\crefname{section}{Sec.\!}{Secs.\!}
\crefname{equation}{Eq.\!}{Eqs.\!}
\crefname{figure}{Fig.\!}{Figs.\!}
\crefname{table}{Tab.\!}{Tabs.\!}
\crefname{appendix}{App.\!}{Apps.\!}

\newcommand{\av}[1]{\langle #1 \rangle}

\begin{document}
\title{Functional Renormalization Group Study
of Thermodynamic Geometry 
Around the Phase Transition of Quantum Chromodynamics}
\author{Fabrizio Murgana}
    
    \affiliation{
        Department of Physics and Astronomy, University of Catania,
        Via S.\  Sofia 64, I-95125 Catania, Italy
    }
    \affiliation{
        INFN-Sezione di Catania,
        Via S.\ Sofia 64, I-95123 Catania, Italy
    }
	\affiliation{
		Institut f\"ur Theoretische Physik, Goethe-Universit\"at,
		Max-von-Laue-Stra{\ss}e 1, D-60438 Frankfurt am Main, Germany
	}

\author{Vincenzo Greco}
    
    \affiliation{
        Department of Physics and Astronomy, University of Catania,
        Via S.\  Sofia 64, I-95125 Catania, Italy
    }
    \affiliation{
        Laboratori Nazionali del Sud, INFN-LNS, Via S. Sofia 62, I-95123 Catania, Italy
    }

\author{Marco Ruggieri}

    
    \affiliation{
        Department of Physics and Astronomy, University of Catania,
        Via S.\  Sofia 64, I-95125 Catania, Italy
    }
    \affiliation{
        INFN-Sezione di Catania,
        Via S.\ Sofia 64, I-95123 Catania, Italy
    }

\author{Dario Zappal\`a}

    \affiliation{
        INFN-Sezione di Catania,
        Via S.\ Sofia 64, I-95123 Catania, Italy
    }
    \affiliation{
        Centro Siciliano di Fisica Nucleare e Struttura delle
        Materia, 
        Via S.\ Sofia 64, I-95123 Catania, Italy
    }


\begin{abstract}
    We investigate the thermodynamic geometry of the quark-meson model 
    at finite temperature, $T$, and quark number chemical potential,
    $\mu$.
We extend previous works by the inclusion of
fluctuations exploiting the functional renormalization group approach.  
We use recent developments 
to recast the flow 
equation into the form of an advection-diffusion equation.
We adopt the local potential approximation for the effective average action.
We focus on the thermodynamic curvature, $R$, in the $(\mu,T)$ plane,
in proximity of the chiral crossover, up to the critical point
of the phase diagram. We find that the inclusion of fluctuations
results in a smoother behavior
of $R$ near the chiral crossover.
Moreover, for small $\mu$, $R$ remains negative, signaling the fact that bosonic fluctuations reduce
the capability of the system to completely overcome the 
fermionic statistical repulsion of the quarks. 
We investigate in more detail the small $\mu$ region
by analyzing
a system in which we artificially lower the pion mass,
thus approaching the chiral limit in which the crossover is
actually a second order phase transition.
On the other hand,
as $\mu$ is increased and the critical point is approached,
we find that $R$ is enhanced and a sign change
occurs, in agreement with mean field studies. Hence, 
we completely support the picture that $R$ is sensitive to
a crossover and a phase transition, 
and provides information about the 
effective behavior of the system at the phase transition.
\end{abstract}

\maketitle

\section{Introduction}

Within the theory of fluctuations
among equilibrium states, one of the most interesting ideas is given by the  
thermodynamic curvature \cite{Ruppeiner:79,Ruppeiner:1981znl,Ruppeiner:83,Ruppeiner:1983zz,Ruppeiner:85a,Ruppeiner:85b,Ruppeiner:86,Ruppeiner:90a,Ruppeiner:90b,Ruppeiner:91,Ruppeiner:93,Ruppeiner:1995zz,Ruppeiner:98,Ruppeiner:05,Ruppeiner:10,Ruppeiner:2008kd,Ruppeiner:2011gm,Ruppeiner:2012uc,Ruppeiner:12,Ruppeiner:13,Ruppeiner:2013yca,Ruppeiner:14,Ruppeiner:15a,Ruppeiner:15b,Ruppeiner:16,Ruppeiner:17}, which represents an innovative perspective in the field of thermodynamics. This employs non-euclidean geometry to represent fluctuations and interactions among thermodynamic variables. The geometric approach sheds new light on understanding phase transitions and emergent properties in complex systems, providing an intriguing connection between geometry and thermodynamics. In fact, in the grand-canonical ensemble one can consider 
any pair of intensive variables
$(\beta^1,  \beta^2)$: given these, the probability of a fluctuation from the state $S_1 = (\beta^1,  \beta^2)$
to $S_2=(\beta^1 + \delta \beta^1,  \beta^2+\delta \beta^2) $  is proportional to
\begin{equation}
\sqrt{g} \exp\left({-\frac{d\ell^2}{2}}\right).
\end{equation}
Here, 
\begin{align}
    d \ell^2= g_{\beta^1 \beta^1} d\beta^1 d\beta^1 + 2 g_{\beta^1 \beta^2} d\beta^1 d\beta^2 + g_{\beta^2 \beta^2} d\beta^2 d\beta^2 , 
    \label{eq:eq1_murgana_intro}
\end{align}
where 
\begin{align}\label{eq:gg_intro}
     g_{ij} =\frac{\partial^2 \log \mathcal{Z}}{ \partial\beta^i \partial\beta^j } 
\end{align}
is the metric tensor in the 2-dimensional manifold; $\mathcal{Z}$ is the grand-canonical partition function.  
Finally, $g$ is the determinant of the metric tensor.  $d \ell^2$ 
measures the distance between the states $S_1$ and $S_2$. Equipped with the metric, we introduce the thermodynamic curvature, $R=2R_{1212}/g$, where $ R_{1212}$ corresponds to the only
independent component of the Riemann tensor for a twodimensional manifold. $R$ depends on the second
and third order moments of the thermodynamic variables 
that are conjugated to $(\beta^1,  \beta^2)$:  it thus carries information about the fluctuation of the physical quantities. 

Within the context of thermodynamic geometry,  interesting and recent  developments regard its applications to effective models, in particular to the quark-meson (QM) model~\cite{Zhang:2019neb, Castorina:2020vbh}. The QM  model is a well known low energy effective model of Quantum Cromodynamics (QCD) \cite{Gell-Mann:60,Koch:97ei, Peskin:95,Weinberg:96}; it has mostly been used because it could qualitatively capture the  chiral phase transition of QCD, i.e the transition from chiral symmetry broken phase in QCD vacuum, to a chiral symmetry
restored phase at finite temperature, $T$, and quark number chemical
potential, $\mu$.  
\begin{figure}[t!]
                \includegraphics[width=\linewidth]{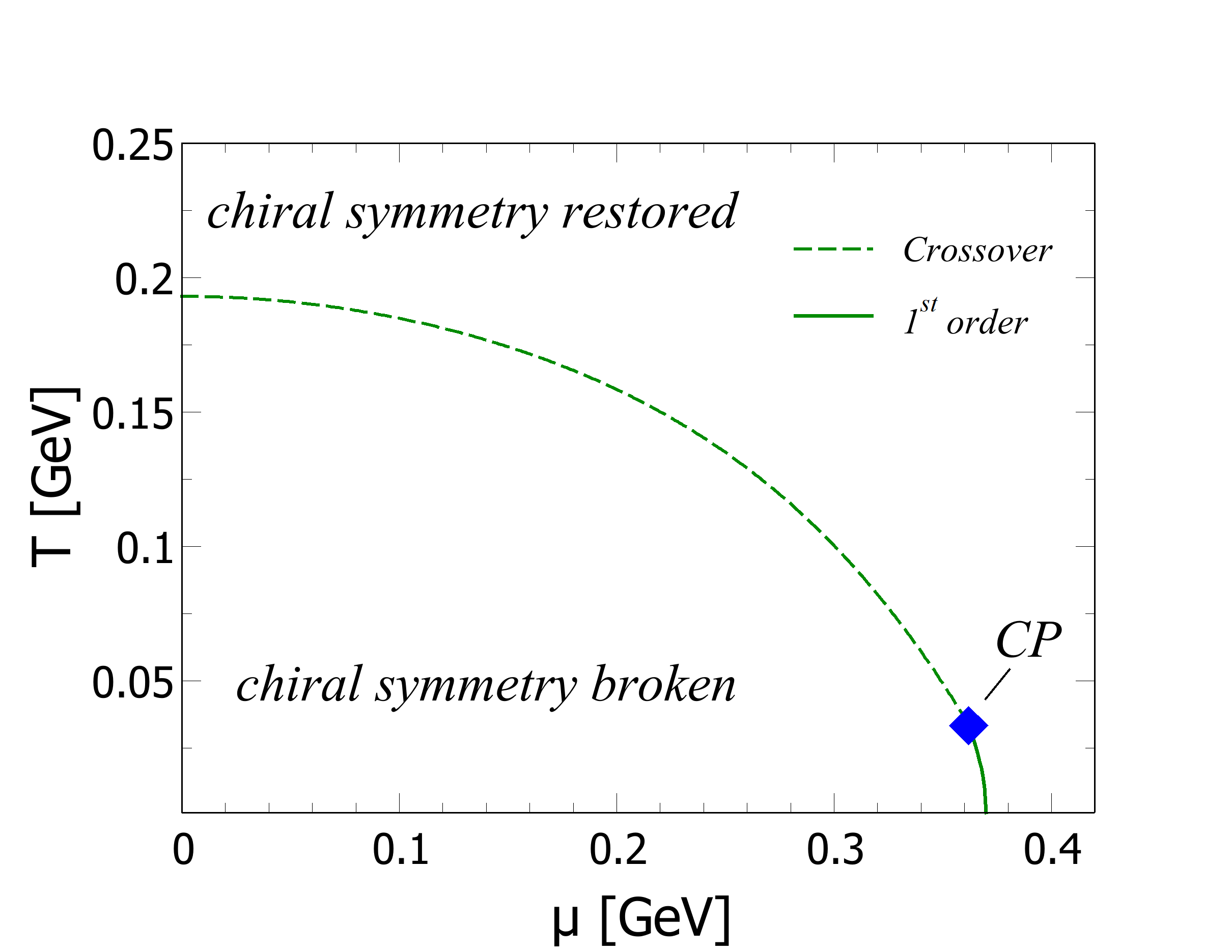}
            \caption{Phase diagram of the quark-meson model at finite
temperature T and chemical potential $\mu$. The chirally spontaneously (and explicitly) broken phase and the (approximately) chirally symmetric phase are indicated in the plot. The dashed line, indicating a crossover, and the 1st order phase transition lines merge at the critical end point, labeled CP.}
            \label{Fig:pdintro}
\end{figure}
In Fig.~\ref{Fig:pdintro} we show a cartoon phase diagram
of the QM model in the $(\mu,T)$ plane; 
the diagram is drawn assuming a finite quark mass which 
explicitly breaks
the chiral symmetry.
In the figure,
the dashed line corresponds to a smooth crossover, while the
solid line to a first order transition; in both cases,
chiral symmetry is spontaneously broken in the low temperature
phase while it is (approximately) restored in the high
temperature phase. 
The two lines meet at a point, known as the
critical endpoint (CP). At this point, the crossover
becomes a real second order phase transition
with a divergent correlation length. 
The purpose of the present work is to study the
thermodynamic curvature around the crossover,
both at small $\mu$ and in the proximity of CP.

One of the most viable uses of the QM model
is related to the study of the strongly interacting 
matter at finite $T$ and/or $\mu$. 
Similarly to the already discussed phase diagram of the QM model,
first principle calculations performed in Lattice-QCD (LQCD)
\cite{Cheng:10,Bazavov:12,Borsanyi:10}
show that at $\mu=0$, 
a smooth crossover happens between a low-temperature phase,
in which chiral symmetry is spontaneously broken and the relevant
degrees of freedom are hadrons, to a quark-gluon plasma phase 
at high temperature
where chiral symmetry
is approximately restored and the relevant degrees of freedom
are quarks and gluons. 
In addition to this,
it is commonly believed that  there should exist a critical endpoint in the $(T,\mu)$ plane in full QCD, 
similar to the one found in the QM model,
see~\cite{Fukushima_2011, Bzdak_2020} for reviews.
However, first principle calculations are not
feasible at finite $\mu$
due to the (in)famous sign problem \cite{Cheng:10,Bazavov:12,Borsanyi:10}. 
This is where
effective models, like the QM model, can help.

In this work, we study the QM model at finite $T$ and $\mu$, focusing of its thermodynamic geometry;
this is the natural continuation of previous works on
the same subject~\cite{Zhang:2019neb, Castorina:2020vbh,Castorina:2019jzw}. 
The aspect of novelty of the study is the inclusion 
of the quantum fluctuations via the functional renormalization group approach (FRG)~\cite{Pawlowski:2005xe,Gies:12,Kopietz:2010zz,Dupuis:2020fhh,Berges:02,Wilson:1971dh,Wilson:1971bg,WILSON:75,Wilson:1974mb,Wilson:1979qg,Polchinski:1983gv,Hasenfratz:1985dm,Ellwanger:1993mw,Wetterich:1992yh,Morris:94,Reuter:1993kw,WETTERICH:91,Morris:1993qb,Koenigstein:2021syz,Koenigstein:2021rxj,Koenigstein:21III,Bergerhoff:98,Tetradis:1995br,Berges:95}; 
this is based on the ideas introduced by Wilson \cite{Wilson:1971dh,Wilson:1971bg,WILSON:75,Wilson:1974mb,Wilson:1979qg} and others, see, e.g., Refs.~\cite{Polchinski:1983gv,Hasenfratz:1985dm}. Since the FRG method belongs to the class of non-perturbative approaches to quantum field theories and it is built in such a way to link  different energy scales and the associated degrees of freedom, it is a suitable tool to deal with second order phase transitions and critical long wave length phenomena in general, whose nature is highly non-perturbative.

The reason why the investigation of the critical behaviour of the QM model via the thermodynamic geometry is worthy lies in the information one can extract from $R$ on the system. Particularly,
it has been argued that the sign of $R$ is connected to the dominance of a fermionic or bosonic-like behavior of the system~\cite{Ruppeiner:05,Ruppeiner:10,Ruppeiner:90b,Ruppeiner:07}:
a positive $R$ corresponds to a boson-like behavior while
a negative $R$ to a fermion-like behavior. 
This behavior has been understood as
attraction or repulsion in the phase space~\cite{Ruppeiner:05,Ruppeiner:10,Ruppeiner:90b,Ruppeiner:07}.
Therefore,
the study of the sign of $R$ near a second order phase transition
(and to some extent, around a crossover~\cite{Zhang:2019neb, Castorina:2020vbh,Castorina:2019jzw}) 
can reveal information about the effective dynamics
which develops in the proximity of this transition.
Another interesting feature of $R$,
anticipated in~\cite{Zhang:2019neb, Castorina:2020vbh,Castorina:2019jzw},
is linked to the presence of a peak structure close to the crossover temperature for small values of $\mu$, which eventually should turn into a divergence when the critical point is reached. 
We complete the studies of~\cite{Zhang:2019neb, Castorina:2020vbh,Castorina:2019jzw}
by analyzing these aspects of $R$ within the FRG approach.

This article is structured as follows: In Sec. II, we
briefly introduce the thermodynamic geometry. 
In Sec. III, we discuss the
basic features of the quark-meson model. 
In Sec. IV,  we present a short introduction to the
functional renormalization group
and how this can be
specifically
applied to the  
quark-meson model. 
In Sec. V, we discuss the results  on the 
thermodynamic geometry for the quark-meson model. 
Finally,
in Sec. VI we 
draw our conclusions and present
an  outlook. 
Within this article we use natural units
$\hbar=c=k_B=1$.

\section{Thermodynamic Curvature}
We consider a thermodynamic system in the grand-canonical
ensemble whose equilibrium state is characterized by the
pair $(\mu, T)$, where $T$ is the temperature and $\mu$ is the 
quark number chemical
potential (conjugated to 
quark number
density). Thermodynamic geometry is 
more conveniently defined in terms
of the variables $(\beta=1/T\, ,\,  \gamma=-\mu/T)$.
A thermodynamic system at
equilibrium 
at the point $(\beta,\gamma)$
can fluctuate to another equilibrium state
$(\beta^\prime,\gamma^\prime)$, 
and the probability
of this fluctuation can be computed within the standard
thermodynamic fluctuation theory. 
In fact, as already mentioned in the introduction, 
we firstly define 
a distance
in the two-dimensional
manifold spanned by $(\beta,\gamma)$,
\begin{align}
    d \ell^2= g_{\beta \beta} d\beta d\beta + 2 g_{\beta \gamma} d\beta d\gamma + g_{\gamma \gamma} d\gamma d\gamma 
    \label{eq:eq1_murgana}
\end{align}
where the metric tensor is 
\begin{align}\label{eq:gg}
     g_{ij} =\frac{\partial^2 \log \mathcal{Z}}{ \partial\beta^i \partial\beta^j } = \frac{\partial^2 \phi}{ \partial\beta^i \partial\beta^j } \equiv \phi_{,ij},
\end{align}
with $\phi=\beta P$, $P=-\Omega$ and $\Omega$ 
denotes 
the thermodynamic potential density; 
moreover, we used the standard notation
$\beta^1=\beta$ and $\beta^2=\gamma$.
Given these, the fluctuation probability is
\begin{equation}
\frac{dp}{d\beta d\gamma}
\propto \sqrt{g} \exp\left({-\frac{d\ell^2}{2}}\right).
\label{eq:fluct_giada}
\end{equation}
where
\begin{align}
     g = g_{\beta \beta}  g_{\gamma \gamma} -  g_{\beta \gamma}^2
\end{align}
is the determinant of the metric. 
Large probability of a fluctuation corresponds to
small $d\ell^2$. Therefore,
a large thermodynamic distance between two equilibrium
states means a small probability to
fluctuate between the two states. According to these considerations, Eq.~\eqref{eq:eq1_murgana}
measures the distance in the $(\beta,\gamma)$  
plane between two
thermodynamic states in equilibrium.

Thermodynamic stability requires that $g_{\beta\beta} > 0$ and
$g > 0$, while $g = 0$ corresponds to a phase boundary and
regions with $g < 0$ are thermodynamically 
unstable: hence, the stability conditions
ensure that $d\ell^2 >0$. 
Furthermore, 
\begin{eqnarray}
&&V\phi_{,\beta \beta}= \av{(U-\av{U})^2},\\
&&V\phi_{,\beta \gamma}= \av{(U-\av{U})}\av{(N-\av{N})},\\
&& V\phi_{,\gamma \gamma}= \av{(N-\av{N})^2}.
\end{eqnarray}
where $U$ and $N$ denote the internal energy and the particle
number, respectively, while $V$ stands for the volume
of the system.

Once the manifold has been  provided with the metric tensor, 
one can define the Riemann tensor as
\begin{equation}
    R^i_{klm}=\frac{\partial \Gamma^i_{km}}{\partial x^l}-\frac{\partial \Gamma^i_{kl}}{\partial x^m}+\Gamma^i_{nl}\Gamma^n_{km}- \Gamma^i_{nm}\Gamma^n_{kl},
\end{equation}
with the Christoffel symbols
\begin{equation}
    \Gamma^i_{kl}=\frac{1}{2}g^{im}\left(\frac{\partial g_{mk}}{\partial x^l}+\frac{\partial g_{il}}{\partial x^k}-\frac{\partial g_{kl}}{\partial x^m}\right).
\end{equation}
The standard contraction procedure gives the
Ricci tensor $R_{ij}= R^k_{
ikj}$, and the scalar curvature $R=R^i_
i$:
within thermodynamic geometry,
$R$ is called the thermodynamic curvature.
For the two-dimensional manifold 
that we consider in this study
the expression of $R$ considerably
simplifies, namely~\cite{Ruppeiner:1995zz}
\begin{align}
    R=-\frac{1}{2\, g^2} \begin{vmatrix}
        \phi_{,\beta\beta} & \phi_{,\beta\gamma} & 
        \phi_{,\gamma \gamma}\\
        \phi_{,\beta\beta\beta} & \phi_{,\beta\beta\gamma} & \phi_{,\beta\gamma\gamma}\\
        \phi_{,\beta\beta\gamma} & \phi_{,\beta\gamma\gamma} & \phi_{,\gamma\gamma\gamma}
        \end{vmatrix},
\label{eq:R_2d_ggg}
\end{align}
where $||$ indicates the determinant of the matrix.
The curvature diverges for $g \rightarrow 0$,  which corresponds to a phase
boundary, unless the numerator of
Eq.~\eqref{eq:R_2d_ggg} vanishes on the same boundary.

Let $\xi$ denote the correlation length of the order parameter:
then, $|R|\propto \xi^3$ near a
second-order phase transition~\cite{Ruppeiner:79},\
which naturally results from hyperscaling.
Theoretical calculations based on different
models confirm this hypothesis \cite{Ruppeiner:79,Ruppeiner:1995zz,Janyszek:1989zz,Janyszek:90,Murgala:92}; therefore,
the study of $R$ in the $(\mu,T)$  plane allows to
estimate the correlation volume based only on the
thermodynamic potential:
this is one of the merits of the thermodynamic
geometry.
It has also been suggested that the sign of $R$ conveys
details about the nature of the interaction, attractive or
repulsive, at a mesoscopic level in proximity of the phase
transition.

Within our sign convention, $R > 0$ indicates an
attractive interaction while $R < 0$ corresponds to a repulsive
one. These interactions include not only real interactions \cite{Ruppeiner:12,Ruppeiner:15b,Ruppeiner:2011gm,May:12,May:13}, but also the statistical attraction and
repulsion that ideal quantum gases feel in phase space \cite{Mirza:2008fy,Oshima:99,Ubriaco:13,Ubriaco:16,Mehri:20}:
an ideal fermion gas has $R < 0$ due to the
statistical repulsion, while an
ideal boson gas has $R > 0$ due to the statistical attraction. 
The thermodynamic curvature is known to be
identically zero only for the ideal classical gas. Other fields
of application of thermodynamic geometry include Lennard-Jones fluids~\cite{Ruppeiner:2011gm,May:12,May:13}, ferromagnetic
systems~\cite{Dey:13}, black
holes \cite{Chaturvedi:17,Sahay:2017hlq,Shay:10,Aman:03,Shen:07,Aman:2005xk,Ruppeiner:2011gm,Sarkar:08,Bellucci:12,Wei:13,Wei:16,Sahay:17,Ruppeiner:18,Ruppeiner:07,Yerra:20,Yerra:21,Lanteri:21}, strongly interacting matter~\cite{Castorina:18,Castorina:20,Zhang:2019neb} and
others~\cite{Diosi:87,Diosi:89}.

\section{The Quark-Meson Model}
The Quark-Meson model (QM)  can be understood as arising from the well-known linear-sigma model coupled to fermions (\cite{Gell-Mann:60,Koch:97ei, Peskin:95,Weinberg:96}). In particular, the $N_f=2$ QM model uses as fundamental degrees of freedom four mesons, an isotriplet of pion fields $\vec{\pi}=(\pi^1, \pi^2, \pi^3)$ and an isosinglet field $\sigma$, coupled to a massless isodoublet fermionic field $\psi$ representing up and down quarks. The euclidean lagrangian density of the model then reads
\begin{eqnarray}
\mathcal{L}^E_{QM} &=& \bar{\psi}(\gamma_\mu\partial^\mu+h(\sigma+i\gamma^5\vec{\tau}\vec{\pi}))\psi
\nonumber\\
&&
+\frac{1}{2}(\partial_\mu\sigma)^2+\frac{1}{2}(\partial_\mu\vec{\pi})^2+U(\sigma^2+\vec{\pi}^2)- c\sigma.
\nonumber\\
&&\label{eq:qm_lagr_111}
	\end{eqnarray}
in which $\gamma_\mu$ and $\gamma^5$ are the standard euclidean Dirac matrices in Dirac space, $h$ is the strength of the Yukawa coupling between quarks and mesons and   $\vec{\tau} = (\tau^1, \tau^2, \tau^3)$ represents the vector of  Pauli matrices in flavor space. $U(\sigma^2+\vec{\pi}^2)$ indicates a generic interaction potential among the mesons, and it constructed in order to be O$(4)$ symmetric, since it depends on the O$(4)$ invariant mesonic combination $\sigma^2+\vec{\pi}^2$. Anyway,  if the potential
develops a finite minimum along one radial direction, the  O$(4)$ symmetry is spontaneously broken into a  O$(3)$ symmetry. Due to this residual symmetry, one is free to choose the ground state in the vacuum as 
\begin{equation}
\av{\vec{\pi}}=0,~~~\av{\sigma}=f_\pi\neq 0,
\end{equation}
where $f_\pi=0.093$ GeV is the pion decay constant. 

Even though the lagrangian does not contain an explicit mass term for the fermions, when the symmetry is spontaneously broken they acquire a dynamical constituent mass given by $M=h\av{\sigma}$. Furthermore, the O$(4)$ symmetry is also explicitly broken by the term $-c\sigma$, which mimics the presence of a finite current mass for quarks. In this way, the pions turn into massive pseudo-Goldstone mesons, since the spontaneous symmetry breaking pattern is not exact, acquiring a finite mass $M^2_\pi=c/f_\pi$.

\section{The Functional Renormalization Group}
	  In this section, we provide a brief introduction to the FRG,
following the interpretation provided by Wetterich {\it et al.}~\cite{Ellwanger:1993mw, Wetterich:1992yh,Morris:94,Reuter:1993kw},
see~\cite{Pawlowski:2005xe,Gies:12,Kopietz:2010zz,Dupuis:2020fhh,Berges:02}
for reviews.
    When performing an FRG study, one changes the focus from the usual generating functional of the theory, or the partition function, to the  effective action $\Gamma [ \Phi ]$, which serves as the generating functional of 1PI vertex functions. 
Within the FRG formalism, similarly to Wilson's approach to renormalization,
for the computation of $\Gamma [ \Phi ]$
one integrates out the fluctuations by successive momentum shells;
this is achieved by  introducing the effective average action, 
$\Gamma_k [ \Phi ]$,  which depends on $k$ 
that represents
the momentum up to which fluctuations have been effectively integrated out, 
thus serving as a coarse-graining scale. 
In particular, the effective average action interpolates between the bare classical action, $S_{\textrm{bare}} [ \Phi ]$, 
in the UV, i.e, at the cutoff scale $k=\Lambda$ when no fluctuations have been taken into account, and the full quantum effective action $\Gamma [ \Phi ]$ in the IR, i.e. $k=0$, when all quantum fluctuations have been 
integrated out, namely
\begin{equation}
\Gamma_{k \to \infty} [ \Phi ] = S_{\textrm{bare}} [ \Phi ],~~~\Gamma_{k \to 0} [ \Phi ] = \Gamma [ \Phi ].
\label{eq:occo1}
\end{equation}
The evolution of the effective average action from the UV to the IR
is described by the  FRG flow equation, that is \cite{WETTERICH:91,Wetterich:1992yh,Morris:1993qb}
\begin{equation}\label{eq:3}
\partial_k \Gamma_k [ \Phi ] = \mathrm{Tr} \Big[ \big( \tfrac{1}{2} \,  \partial_k R_k \big) \big( \Gamma^{(2)}_k [ \Phi ] + R_k \big)^{-1} \Big] 
\end{equation}
in which the trace is understood  as the sum over internal degrees of freedom of the theory, such as color, flavor, spin etc, as well as  
an integral over Fourier momenta.
$R_k$ in \cref{eq:3} is a regulator
which effectively defines the support of the momentum
integral in the UV and acts as a screening term in the IR.
$R_k$ 
must be chosen in order for $\Gamma_k$ to fulfill
the conditions~\eqref{eq:occo1}. 
{In particular,} 
in order to recover the full quantum effective action in the IR, it has to vanish for $k \to 0$. Moreover, $R_k$ has to
diverge for large $k$ in order to ensure that the bare action represents a stationary point for the path integral through which $\Gamma_k$ is introduced.

$G_k [ \Phi ] \equiv ( {\Gamma}^{(2)}_k [ \Phi ] + R_k )^{-1}$ in \cref{eq:3} indicates the exact, scale dependent  propagator, thus the FRG flow equation has a one-loop structure. 
Despite this apparent simplicity, it 
still consists of a  hard-to-solve 
functional integro-differential equation; hence, 
except for very special cases
(see for example~\cite{Koenigstein:2021syz,Koenigstein:2021rxj,Koenigstein:21III}), 
one has to rely on approximations in order to solve it. 
A first possibility is called the vertex-expansion~\cite{Morris:94,Bergerhoff:98}, in which $\Gamma_k [ \Phi ]$ is expanded in powers of the fields $\Phi$ around a certain field configuration $\Phi_0$, and the expansion coefficients are the vertex functions $\Gamma_k^{(n)} ( x_1, \ldots, x_n ) = \delta^n \Gamma_k [ \Phi ]/ \delta \Phi ( x_1 ) \cdots \delta \Phi ( x_n ) \big|_{\Phi = \Phi_0}$.  Plugging the vertex expansion into the FRG flow equation, one obtains a system of infinite coupled integro-differential equations which are then truncated  at certain finite order.

Another approximation scheme,
which is the one we adopt in the present work,
is called the derivative expansion~\cite{Berges:02,Berges:95}, which approximates $\Gamma_k [ \Phi ]$ in terms of powers of space-time derivatives  of the fields. 
Also in this case one would obtain an infinite system of coupled equations, one per each operator compatible with the symmetries of the theory, and thus for a practical solution one has to choose a order of space-time derivatives of the fields to which {the action is truncated}. 
One of the main advantages of the derivative expansion is that one does not need to assume an analytic behaviour in field space of the effective action during the flow evolution, which on the other hand is required by the vertex expansion.

\begin{widetext}

In this work, we focus on 
critical phenomena which involve second order phase transitions (in the chiral
limit), crossovers and critical endpoints in the phase diagram.
It is known that, under these conditions, the free energy is not arbitrarily differentiable (see for example \cite{Mussardo:10, Shang-keng:19, Huang:87, Kopietz:2010zz}); furthermore,  {the} effective action can develop discontinuities, or in general singular points, when dealing with such a kind of phenomena \cite{Grossi:2019urj,Grossi:2021ksl,Stoll:2021ori, Murgana:23on}. This suggests us to take advantage of the properties of the derivative expansion, also at the lowest order called local potential approximation (LPA)~\cite{Murgana:23on}.
Particularly,
we adopt the LPA ansatz for the QM model at finite temperature and quark number density
\begin{eqnarray}
\Gamma_k[\bar{\Psi},\Psi,\phi]&=&
\int_0^{1/T} \!\!dx_4\int \!\!d^3x~\left\{
\bar{\psi}\big(\gamma_\mu\partial^\mu+h(\sigma+i\gamma^5\vec{\tau}\vec{\pi})-\mu\gamma_0\big)\psi
+
\frac{1}{2}(\partial_\mu\sigma)^2+\frac{1}{2}(\partial_\mu\vec{\pi})^2+U_k(\sigma^2+\vec{\pi}^2) -c \sigma\right \}.
\nonumber\\
&&\label{eq:anna_1aa}
	\end{eqnarray}
Using the three-dimensional Litim regulator \cite{Litim:2002cf,Litim:2001up} for both fermions and bosons, 
\begin{align}
     R_{k,B}(p) &= (k^2-\bm{p}^2) \Theta(k^2-\bm{p}^2)\\
     	R_{k,F} (p) &=  i \slashed{p} \left(  \sqrt{\frac{k^2}{\bm{p}^2}} - 1\right) \Theta\left(k^2 - \bm{p}^2 \right),
\end{align}
we get the  flow equation for the effective potential, that is  
 \begin{equation}\label{Flow}
\partial_t U_k(\sigma) = 
-\frac{k^5}{12\pi^2} \left\{ \left[ \frac{1}{E_{k,\sigma}} \coth\left(\frac{E_{k,\sigma}}{2T}\right) + \frac{3}{E_{k,\pi}} \coth\left(\frac{E_{k,\pi}}{2T}\right) \right]
 -4N_c \frac{1}{E_\psi}\left[\tanh\left(\frac{E_\psi-\mu}{2 T}\right)+\tanh\left(\frac{E_\psi+\mu}{2 T}\right)\right] \right\},
\end{equation}
with $t=-\ln{k/\Lambda}$ and 
\begin{align}
E_{k,\sigma}= \sqrt{k^2 + \partial^2_{\sigma} U_k(\sigma)},~~~
E_{k,\pi}=\sqrt{k^2 + \frac{\partial_{\sigma} U_k(\sigma)}{\sigma}}.
\end{align}
Moreover, 
\begin{equation}
E_\psi = \sqrt{\bm p^2 + M^2},
\end{equation}
with $M=h\langle\sigma\rangle$ denoting the previously defined constituent quark
mass.

Analogously to e.g. ~\cite{Murgana:23on,koenigstein:21,Grossi:19}, we
introduce the following variables
\begin{equation}
u_k(\sigma) = \partial_{\sigma} U_k(\sigma),~~~u'_k(\sigma) = \partial_{\sigma} u_k(\sigma).
\end{equation}
Taking the derivative of equation (\ref{Flow}) with respect to $\sigma$ we are able to cast the FRG flow equation into an advection-diffusion equation with a source term \cite{Grossi:2021ksl,Stoll:2021ori}, thus obtaining
\begin{equation}\label{flow_sigma}
\partial_t u_k(\sigma) + \partial_{\sigma} f_k(\sigma,u_k(\sigma)) = \partial_{\sigma} g_k(u'_k(\sigma)) + N_c \partial_{\sigma} S_k(\sigma)
\end{equation}
where we defined the advection flux as
\begin{equation}
f_k(\sigma,u_k(\sigma)) =  \frac{k^5}{4\pi^2}\frac{1}{E_{k,\pi}} \coth\left(\frac{E_{k,\pi}}{2T}\right),
\end{equation}
the diffusion flux, namely
\begin{equation}
g_k(u'_k(\sigma)) = - \frac{k^5}{12\pi^2}\frac{1}{E_{k,\sigma}} \coth\left(\frac{E_{k,\sigma}}{2T}\right),	
\end{equation}
and finally the source term
\begin{equation}
    N_c S_k(\sigma)=\frac{N_c \,k^5}{3\pi^2}\frac{1}{E_\psi}\left[\tanh\left(\frac{E_\psi-\mu}{2 T}\right)+\tanh\left(\frac{E_\psi+\mu}{2 T}\right)\right].
\end{equation}
In this framework, the derivative of the potential $u_k(\sigma)$ plays the role of a conserved quantity, in the sense that it satisfies a generalized conservation law \cite{Grossi:2019urj,Koenigstein:2021syz,Ihssen:2023qaq}. Each of the previously defined contributions arises from the various particle species involved in the model. The advection flux, which is responsible of the bulk motion of the conserved quantity $u$, is originated from the pions. Indeed, there is a  factor 3 that appears in \cref{Flow} and multiplying the advection term. Furthermore, as it can be seen from the definition of the energy $E_{k,\pi}$, the mass term for the pions, $u_k(\sigma)/\sigma$, vanishes at the minimum of the effective potential, in agreement with the nature of the pions as Goldstone bosons (since the explicit symmetry breaking term is linear in the sigma field, it does not contribute to the flow equation and is just added to the IR potential). One can verify that the  speed of characteristics $\partial_u f_k(\sigma,u_k(\sigma))$ is positive if $\sigma<0 $ and negative if $\sigma>0 $, implying that the conserved quantity $u_k(\sigma)$ and the minimum of the potential are always transported towards smaller values of $\sigma$ by the advection. 

On the other hand, the one radial sigma mode produces the  diffusion term, which depends on the curvature mass $u'_k(\sigma)$. The diffusion has no specific direction since it depends on the local gradients of the conserved quantity, meaning that smears out peaks and discontinuities. 

 The fermionic loop gives rise to a time and $\sigma$ dependent source term, which we identify as such since it  is independent of the conserved quantity $u_k(\sigma)$. 

In order to compute thermodynamic quantities we need the effective potential, but solving the  flow equation (\ref{flow_sigma}) we obtain its derivative w.r.t. $\sigma$. This means that we need to integrate the solution in $\sigma$, and so the effective potential would be defined up to an arbitrary integration constant, which is $\sigma$-independent but in principle $T$ and $\mu$-dependent. Thus to obtain the correct thermodynamic properties we need to calculate this constant using directly the flow equation (\ref{Flow}), evaluated in a generic point (that for us is $\sigma=0$). 
Thus, together with \cref{flow_sigma} we also solve
 \begin{align}\label{Flow_zero}
\partial_t U_k(0) =  -\frac{k^5}{12\pi^2} \left\{ \left[ \frac{4}{E_{k,\sigma}(\sigma=0)} \coth\left(\frac{E_{k,\sigma}(\sigma=0)}{2T}\right) \right]\nonumber
-4N_c \frac{1}{k}\left[\tanh\left(\frac{k-\mu}{2 T}\right)+\tanh\left(\frac{k+\mu}{2 T}\right)\right] \right\}.
\end{align}

\end{widetext}

\section{Results}

Throughout this section we compare the results obtained within
the mean field approximation, obtained by neglecting the
bosonic fluctuations, with the calculation that takes into
account the fluctuations by solving the full FRG flow equation
for the average effective action. 
In the mean field approximation
we firstly introduce the rescaling
$\sigma \to \sqrt{N_c} \sigma$, $U_k(\sigma) \to N_c U_k(\sigma)$
and $ u_k(\sigma) \to \sqrt{N_c} u_k(\sigma)$, then
we get the rescaled flow equation
\begin{eqnarray}
&&\partial_t u_k(\sigma) + \frac{1}{\sqrt{N_c}}\partial_{u_k} f_k(\sigma,u_k(\sigma)) u'_k(\sigma)\nonumber\\
&&~~~~~~~~~
= \frac{1}{\sqrt{N_c}}\partial_{\sigma} g_k(u'_k(\sigma)) + \partial_\sigma S_k(\sigma).
	\end{eqnarray}
The mean field flow equation is obtained in the 
$N_c \to \infty$ limit, that is
	\begin{equation}
		\partial_t u_k(\sigma) = \partial_\sigma S_k(\sigma).
	\end{equation}

 As initial condition in the UV ($k=\Lambda$ or $t=0$) for the flow equation we choose a quartic potential
	\begin{equation}
	U_{\Lambda}(\sigma)= \frac{m^2_{UV}}{2}\sigma^2+\frac{\lambda_{UV}}{4}\sigma^4.
	\end{equation}
Due to the presence of a finite cutoff $\Lambda$,  when performing the Matsubara sum,  thermal modes with $2\pi T > \Lambda$ are factually excluded. This {is} a serious problem especially for the calculation of  thermodynamic quantities at high temperatures.  This issue can be fixed including the  missing high-momentum modes in
the effective potential. Since one expects the fermionic degrees of freedom to be relevant at higher temperature, a standard procedure consists of integrating the fermionic part of Eq. (\ref{Flow}) from $k \to \infty$ to $k=\Lambda$ and add it to the effective potential. So we calculate
\begin{equation}
    U^\infty_\Lambda(\sigma)=\int_\infty^\Lambda \, \, S_k(\sigma) dk
\end{equation}
and then add it to the effective potential at the UV scale $\Lambda$
\begin{equation}
    U_\Lambda(\sigma)\, \to \, U_\Lambda(\sigma)+ U^\infty_\Lambda(\sigma)
\end{equation}

Our set of parameters is $\Lambda=1.0$ GeV,  $f_\pi=0.093$ GeV, $h=3.6$, $c=1.78\times 10^{-3}$ GeV$^3$. The initial condition parameters are  $m_{UV}=0.762$ GeV and $\lambda_{UV}=1.05$ for MF calculations,  and $m_{UV}=0.812$ GeV and $\lambda_{UV}=3.08$ in the FRG case. 
These parameters are chosen such that in the vacuum we get $\langle\sigma\rangle=f_\pi$ and $\partial^2_{\sigma\sigma}U(\langle\sigma\rangle)=M^2_\sigma=0.36$  GeV$^2$.

 \begin{figure}[t!]
            
\centering
    \includegraphics[width=\linewidth]{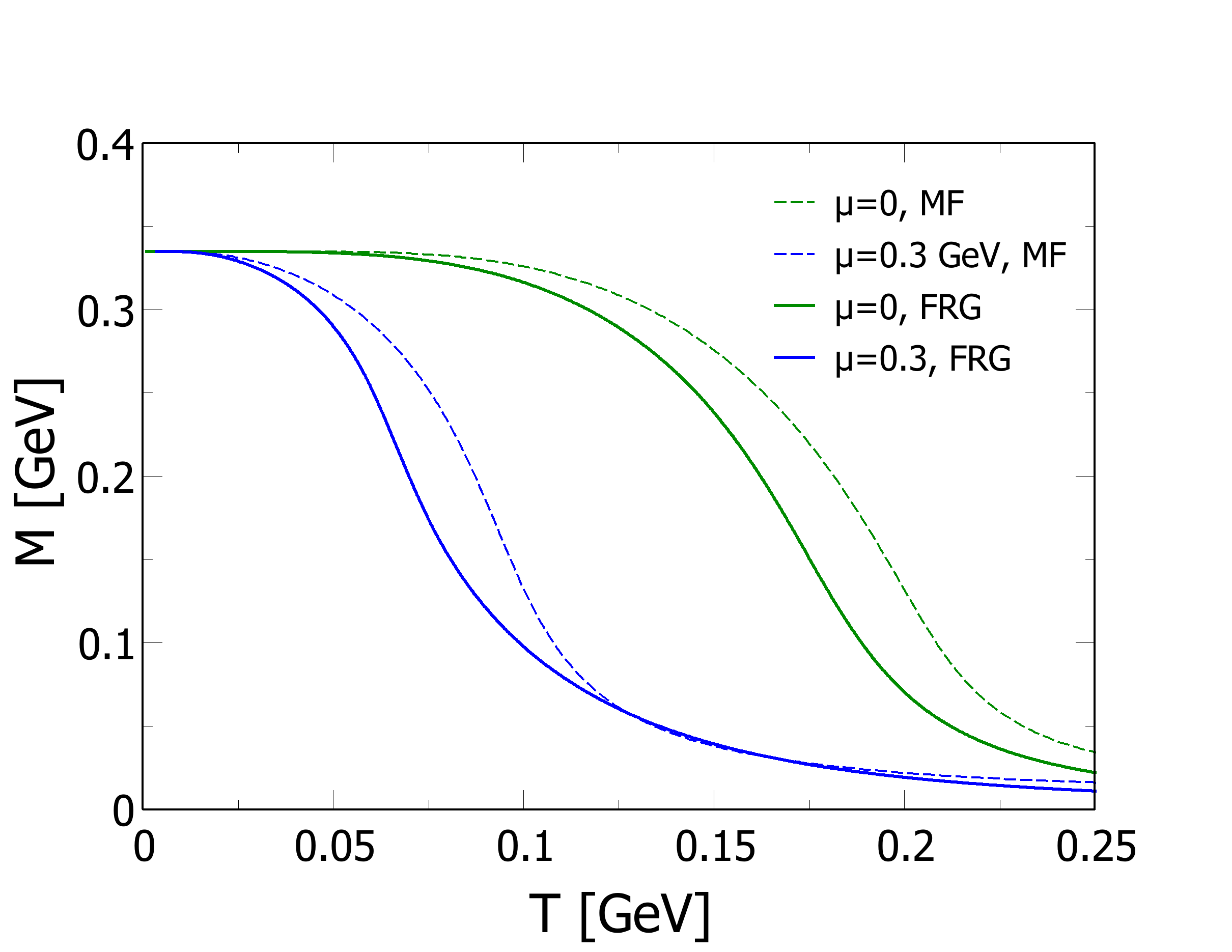}
    \caption{Constituent quark mass,
    $M=h\langle\sigma\rangle$,
    versus temperature, for $\mu=0$ and $\mu=0.3$ GeV.
    MF and FRG stand for mean field and functional renormalization
    group respectively.
    }
            \label{fig:5}
        \end{figure}

In Fig.~\ref{fig:5} we show the  constituent  quark
mass, $M=h\langle\sigma\rangle$,
versus temperature, for $\mu=0$ and $\mu=0.3$ GeV;
$\langle\sigma\rangle$ 
corresponds to the value of $\sigma$ that minimizes the
effective action. 
We note that there is a range of temperatures
in which 
the condensate 
decreases from its zero temperature value to a smaller one,
signaling the crossover from the low temperature phase in which
chiral symmetry is spontaneously broken to the high temperature
phase in which the symmetry is (approximately) restored. 
The picture remains qualitatively the same also after
fluctuations are included; quantitatively,
fluctuations lower the temperature range in which 
the chiral crossover takes place.
We also note that increasing the chemical potential results
in the hardening of the crossover, since changes in 
$\langle\sigma\rangle$ occur in a smaller range of temperature.

\begin{figure}[t!]
            
         \centering
       \includegraphics[width=\linewidth]{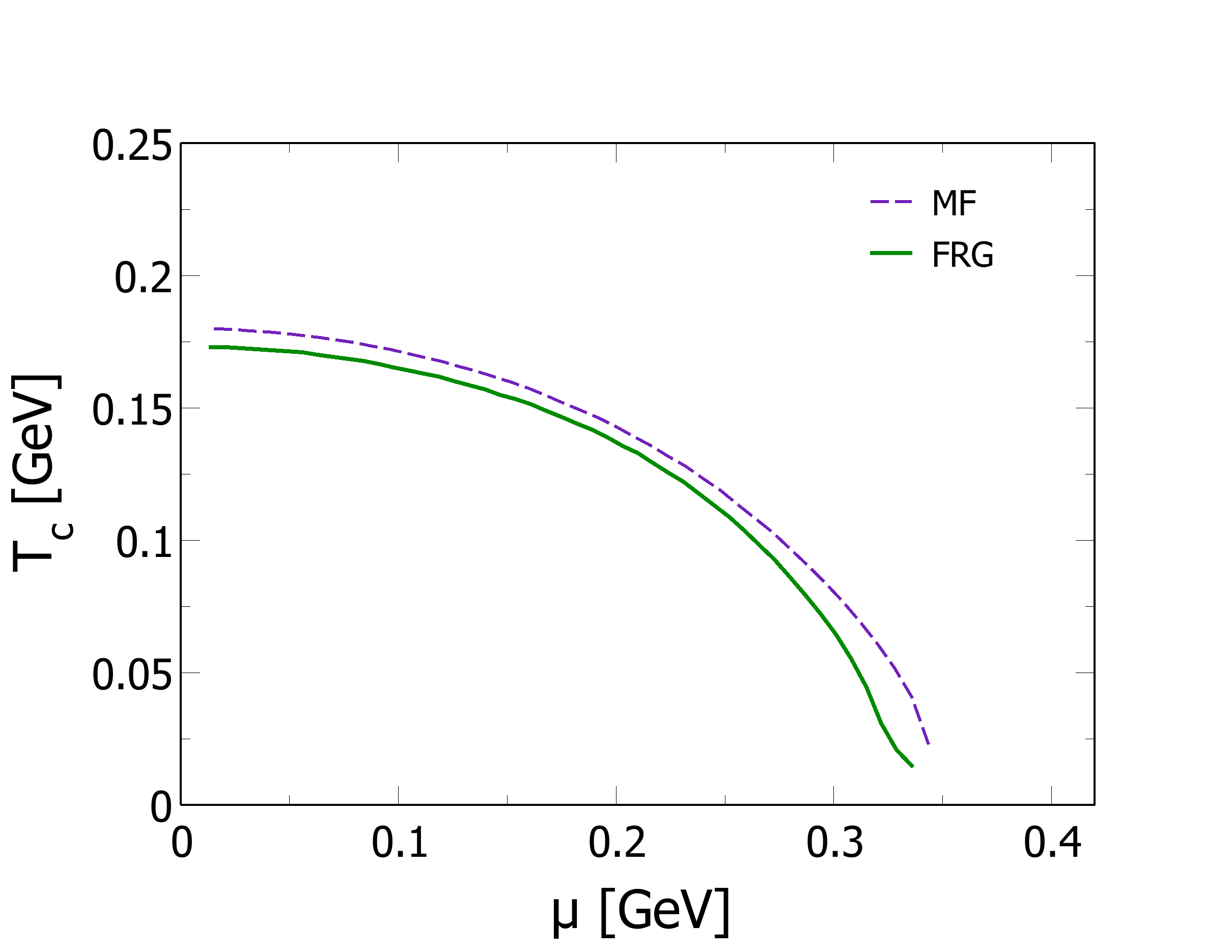}
            
            \caption{\label{Fig:tcmffrg} 
            Critical temperature, $T_c$, versus $\mu$,
            for the mean field (dashed line) and the full FRG
            (solid line) calculations.
            }
          
        \end{figure}

The results shown in Fig.~\ref{fig:5} allow us to define
a (pseudo-)critical temperature, $T_c$, as the temperature at which
the highest change of $\langle\sigma\rangle$ occurs. 
In Fig.~\ref{Fig:tcmffrg} we plot $T_c$ versus $\mu$
for both the mean field case and the full FRG calculation.
Below the critical lines the chiral symmetry is spontaneously broken,
while above the lines chiral symmetry is approximately restored. 
For both cases the lines are stopped at the critical endpoint,
where the crossover changes into a real second order phase
transition  where the chiral susceptibility diverges; for larger values of $\mu$ the phase transition
is of the first order:  in this case the effective potential exhibits two separate finite minima which become degenerate at the phase transition.

\begin{figure}[t!]
            \includegraphics[width=\linewidth]{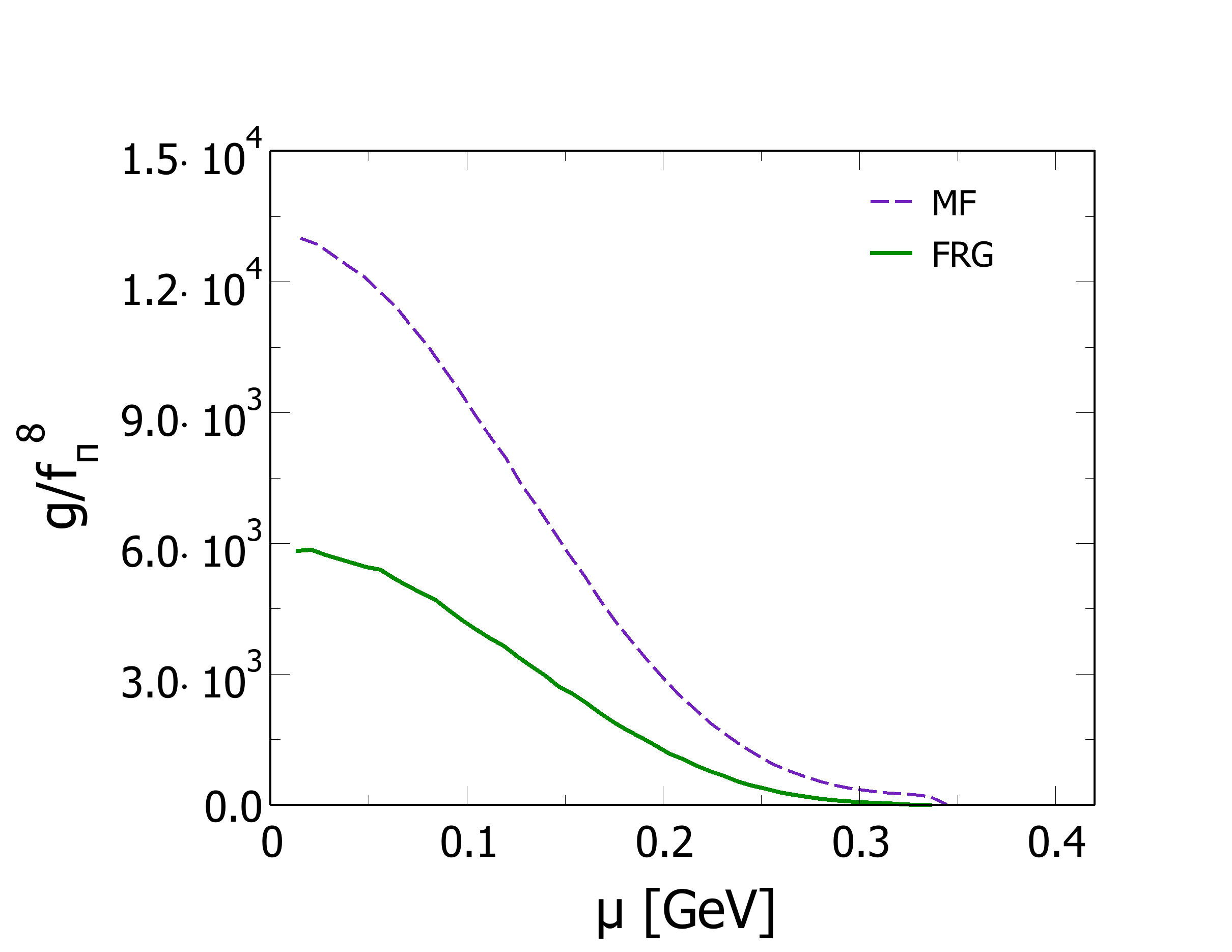}
          \caption{Determinant of the thermodynamic
          metric, $g$, versus $\mu$ computed at $T=T_c$, 
    within the mean field approximation (dashed line)
    and FRG (solid line).
          \label{Fig:determinante}}
            \end{figure}

Next we turn to the main focus of this work, namely
the thermodynamic geometry. Firstly, we show
in Fig.~\ref{Fig:determinante} 
the determinant of the thermodynamic metric,
$g$, versus $\mu$, computed at $T=T_c$.
We present the results obtained within the mean field approximation
and within the FRG.
The results are in qualitative agreement with
\cite{Castorina:2020vbh}, where fluctuations were introduced within a
gaussian approximation. 
$g=0$ corresponds to the thermodynamic instability
of the system, that is to a phase transition. 
We do not find $g=0$ because in this model
chiral symmetry is explicitly, albeit softly, broken by 
the finite quark mass, hence a phase transition is replaced
by a smooth crossover. 
However, $g(T_c)$ decreases with $\mu$,
signaling that the system is approaching criticality,
that is the critical endpoint.
Moreover, we note that
including fluctuations results in the lowering of $g$,
in agreement with previous findings \cite{Castorina:2020vbh}.

\begin{figure}[t!]
            \includegraphics[width=\linewidth]{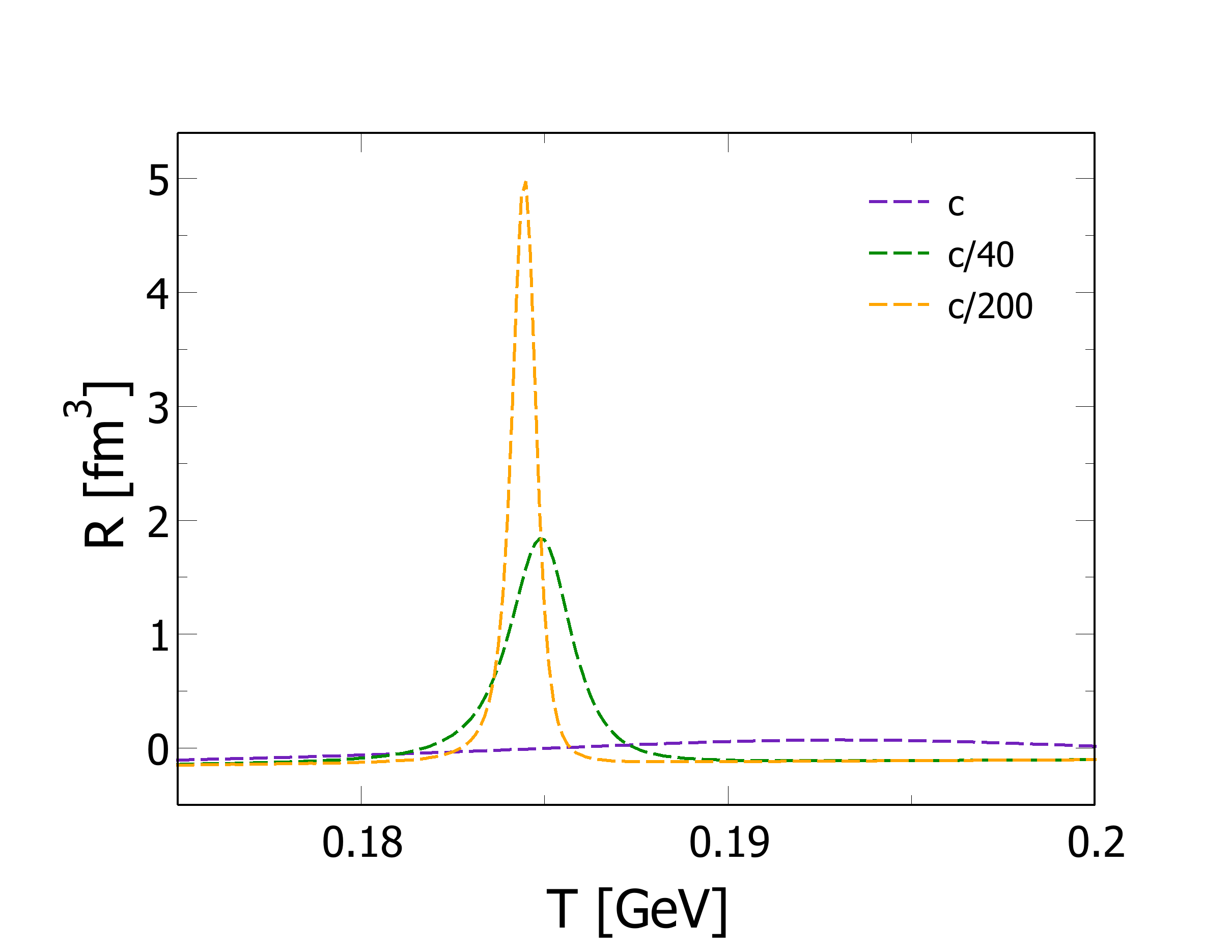}\\
             \includegraphics[width=\linewidth]{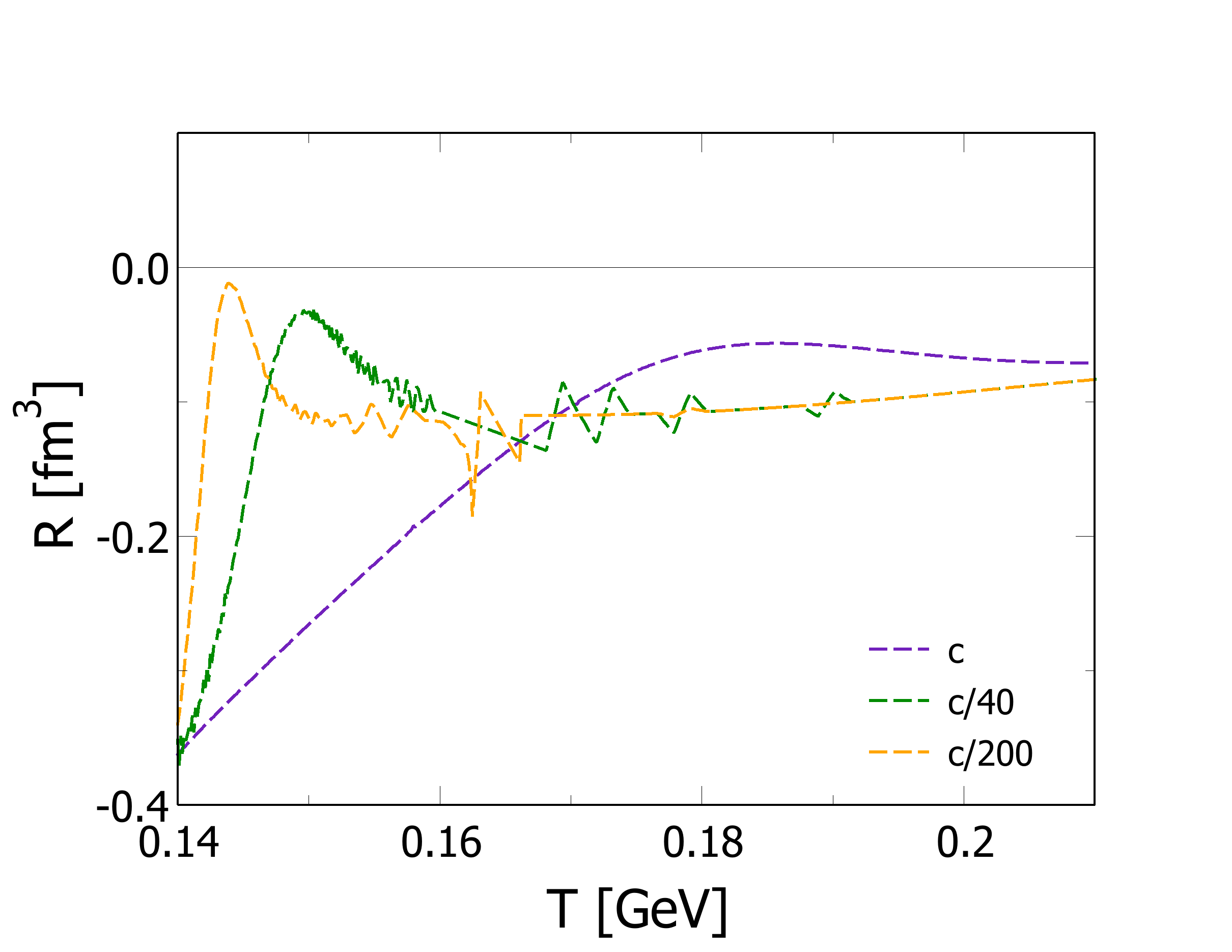}
              \caption{Thermodynamic
              curvature, $R$,
versus $T$, for several values of $c$.              
Upper panel corresponds to the mean field approximation,
while the lower panel to the calculations within the FRG.
$c=M_\pi^2 f_\pi$ denotes the physical value of the symmetry
breaking parameter in Eq.~\eqref{eq:qm_lagr_111}.
              \label{Fig:curvachirale}}
\end{figure}

We now discuss the thermodynamic curvature, $R$.
It is expected that $R$ diverges
at a second order phase transition,
while it is not obvious the behavior of $R$ near a smooth
crossover. 
In order to better understand the results on $R$
obtained within the FRG, we preliminarily
analyze the curvature versus $T$ at $\mu=0$,
with and without fluctuations, for several values of
the parameter $c$ that regulates the explicit breaking of
chiral symmetry.
In fact, in the limit $c=0$, chiral symmetry is not explicitly
broken and restoration of chiral symmetry
at $\mu=0$ happens by a second order phase transition.
Performing calculations of $R$ in the chiral limit
is numerically
demanding near the phase transition, hence we limit
ourselves to analyze cases in which $c$ is small but nonzero.
In order to avoid confusion, from now on with
$c$ we denote solely the value of the parameter at the
physical point, namely $c=M_\pi^2 f_\pi$;
we then artificially lower the value of this parameter.

In Fig.~\ref{Fig:curvachirale}, we plot $R$ versus $T$ at $\mu=0$
within the mean field approximation (upper panel) and
with fluctuations included (lower panel); 
we show results for
several values of the symmetry breaking
parameter in Eq.~\eqref{eq:qm_lagr_111}, 
namely the physical value, then $c/40$
and $c/200$. 
We note in both panels of Fig.~\ref{Fig:curvachirale} that
as the chiral limit is approached, the curvature 
is enhanced in the pseudo-critical region, while the peak of $R$
becomes smoother as $c$ approaches  the one in the physical limit.
Moreover, including fluctuations results in the lowering of the
peaks of $R$ in the pseudo-critical region.
Within the mean field approximation, $R$ changes sign around $T_c$,
in agreement with previous results \cite{Castorina:2019jzw,Castorina:2020vbh,Zhang:2019neb}:
this was interpreted as the emergence of a boson-like behavior
of the system around $T_c$, namely of a
statistical attraction in phase space due to long
range correlations that develop around $T_c$ that overcomes the
statistical fermionic repulsion.
This behavior of $R$ is partially 
found when fluctuations are included
(lower panel of Fig.~\ref{Fig:curvachirale}).
However, we note that fluctuations lower the overall magnitude
of $R$, in agreement with \cite{Castorina:2020vbh}.
We also note that despite the peaks of $R$ become more
prominent as the chiral limit is approached,
it does not change sign as it does in the mean field calculations.
Hence, it is likely that 
although the qualitative behavior of $R$ is independent
on the approximation used (mean field versus FRG),
at the physical point,
the change of nature from fermion-like to boson-like 
depends on the calculation scheme adopted
in proximity of the crossover.
The results shown in Fig.~\ref{Fig:curvachirale} 
will be useful to interpret the behavior of
$R$ we discuss below. As a final remark, we note 
that the peak of $R$ moves towards smaller temperatures as the chiral limit is approached. This is in agreement with the fact that the critical temperature is lower in the chiral limit.
We will see that as the critical point is approached,
the change of sign of $R$ appears both in the mean field and 
in the FRG cases.

\begin{figure}[t!]
\includegraphics[width=\linewidth]{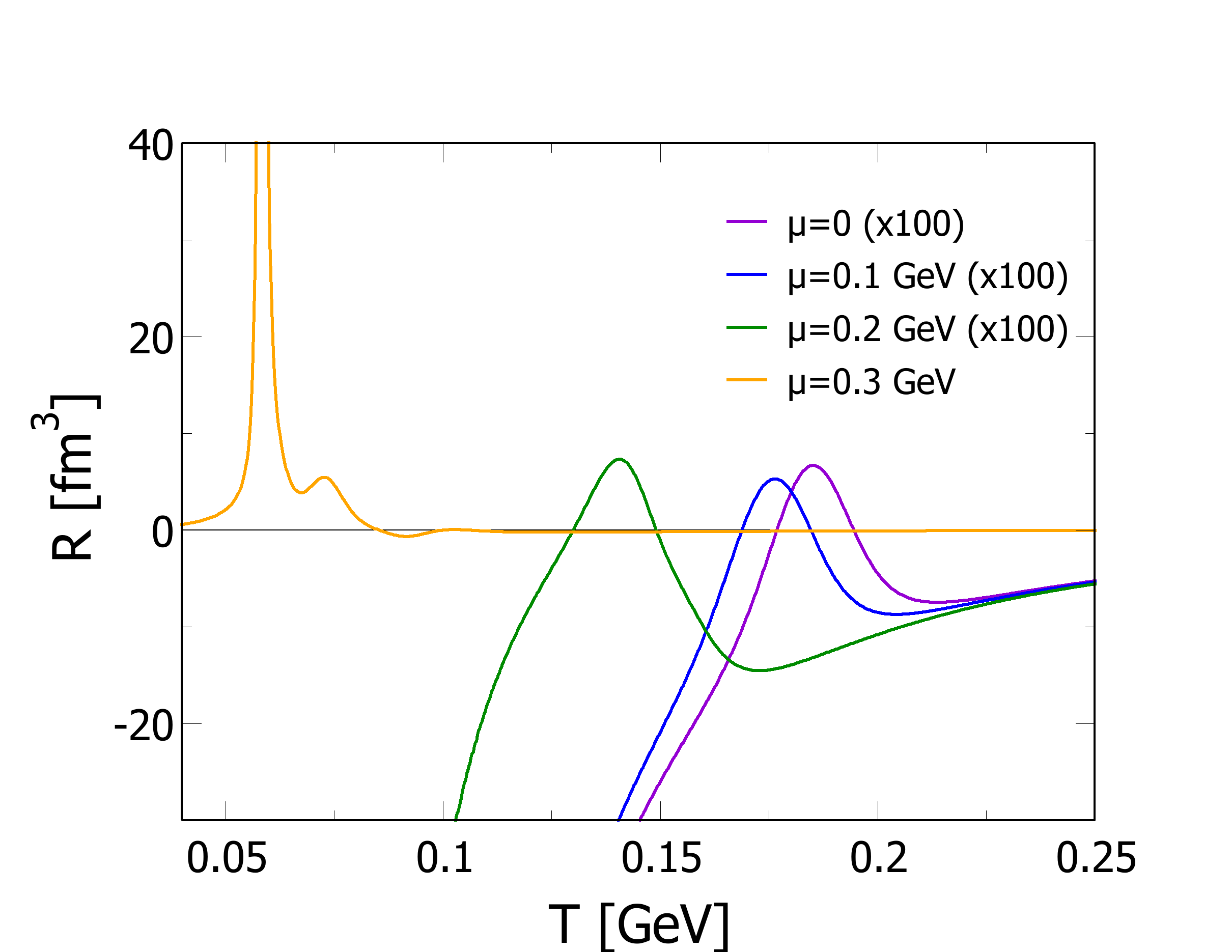}\\
\includegraphics[width=\linewidth]{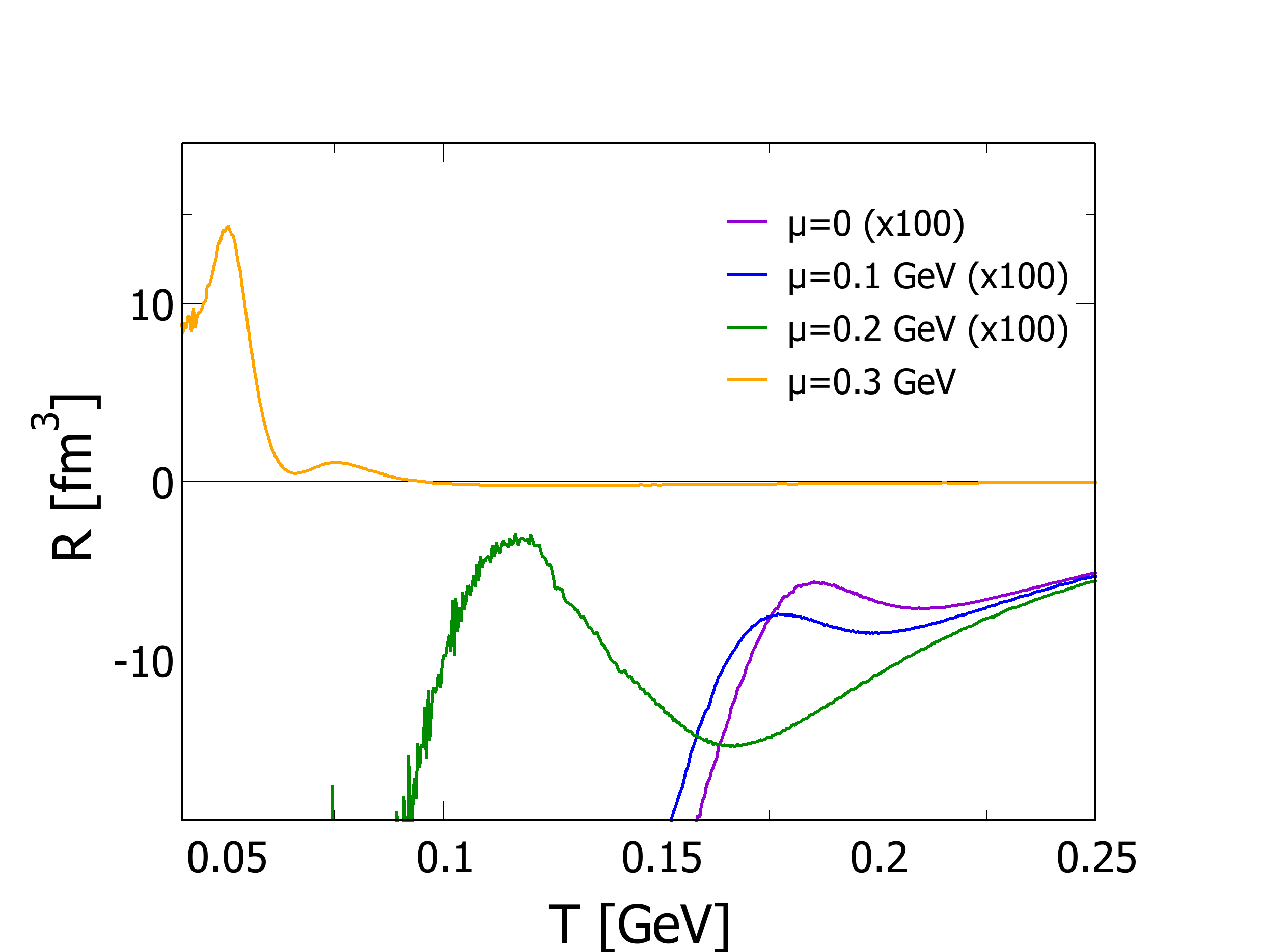}
\caption{Thermodynamic curvature, $R$,
versus $T$, for several values of $\mu$.
Calculations correspond to the mean field
approximation (upper panel) 
and the FRG scheme (lower panel).\label{Fig:curvamffrg}}
\end{figure}

In Fig.~\ref{Fig:curvamffrg} we plot $R$ versus $T$ for several
values of $\mu$, obtained within the mean field approximation
and the full FRG calculation  at the physical 
point  $c=M_\pi^2 f_\pi$. 
In both the MF and FRG panels of Fig.~\ref{Fig:curvamffrg}, we multiplied 
the results by
$100$ in all but the $\mu=0.3$ GeV cases
to make the results more readable. 
Firstly, we note that the trend of $R$ is qualitatively similar
in both calculations. 
Within the mean field approximation, at $\mu=0$ 
the curvature locally develops a peak in correspondence
of the chiral crossover, signaling that $R$ is capable to
capture the pseudo-critical behavior of the quark condensate.
Increasing the chemical potential,
$R$ maintains its local peak structure, but as the critical 
endpoint is approached, the peaks become more pronounced.
This is in agreement with the general understanding on $R$
which is expected to diverge at a second order phase
transition.
Moreover, we note that at large $\mu$ the 
thermodynamic curvature develops several peaks
in the temperature range of the chiral crossover,
although the most pronounced peak does not 
necessarily show up at the
critical temperature. This behavior,
already noticed in \cite{Zhang:2019neb},
shows that $R$ is not necessarily 
as sensitive as other quantities,
like the chiral susceptibility or $|dM/dT|$, at 
the changes of the quark condensate at $T=T_c$,
but it is still capable to measure sensible deviations
in the pressure around the chiral crossover.

Including fluctuations does not change the qualitative behavior
of $R$. Therefore, we conclude that the fact that $R$ is sensitive
to the chiral crossover is 
not an artifact of the mean field approximation, rather it is a 
quite solid statement.
However, as already remarked in Fig.~\ref{Fig:curvachirale},
the inclusion of fluctuations lowers the value of $R$
around the chiral crossover; particularly,
when $\mu$ is small, $R$ remains negative also around the crossover,
while in the mean field approximation it changes sign. 
Therefore, it is likely that the change of nature of the interaction
at the mesoscopic level, from fermion-like to boson-like,
depends on the approximation used in the calculation
when the system is far from criticality. Hence, this implies that, at small $\mu$, the fluctuations substantially change the geometry of the manifold. 

On the other hand, when the critical endpoint is approached
at large $\mu$, we find that $R$ changes sign also in the 
FRG calculation, and the
mean field results do not qualitatively
differ from the those obtained within the FRG. 
Hence, we conclude that when this system
approaches criticality, $R$ changes sign and locally develops
a marked peak: this conclusion was anticipated in previous
mean field calculations \cite{Castorina:2019jzw,Zhang:2019neb}
and stands also in case fluctuations are taken into account
via FRG.

\section{Conclusions and outlook}

We studied the thermodynamic geometry,
and in particular computed the thermodynamic curvature $R$,
of the chiral phase
transition of Quantum Chromodynamics, within
the quark-meson model and the functional renormalization group
method. The advantage of this method is that it allows to 
exactly include fluctuations, 
differently from previous approaches \cite{Castorina:2020vbh}
in which fluctuations were introduced only within a gaussian
approximation.
As a matter of fact, the inclusion of quantum fluctuations via the functional renormalization group represents a remarkable improvement compared to previous mean field calculations. 
We found that the qualitative behavior of $R$ is not very
different from the one previously computed within 
mean field calculations, as well as within
calculation schemes that include fluctuations by a gaussian
approximation. In particular, $R$ seems 
to keep its local peak structure
in proximity of the chiral crossover at small 
chemical potential; moreover, it is enhanced 
at the critical point, signaling that when the system approaches
criticality $R$ could diverge, supporting the 
arguments of hyperscaling~\cite{Ruppeiner:79}.

We also found that the change of sign of $R$ near the
chiral crossover, discussed previously in the literature~\cite{Zhang:2019neb,Castorina:18,Castorina:20,Castorina:2019jzw,Castorina:2020vbh}, does not  always  happen when
fluctuations are taken into account within the functional
renormalization group approach; however, 
as the system approaches criticality, the change of sign
from negative (fermion-like behavior) to positive
(boson-like behavior) takes place. Hence, we 
conclude that
the change of sign of $R$ near the critical endpoint
seems to be quite a robust prediction of the chiral 
effective models.

It will be interesting to analyze 
if the behavior of $R$ we highlighted in this article
does not change when the truncation adopted in the
functional renormalization group approach is improved;
for example, the inclusion of the scale-dependent
wave function renormalization factors of the boson and the quark
fields is worth of further investigation, due to the possible link
of this to the formation of inhomogeneous phases at large
chemical potential. Another possible improvement is the
introduction of other condensation channels,
which could include diquarks or meson condensates.  
A potential
application 
would be
the QM model with a finite isospin chemical potential, 
$\mu_I$,
at vanishing (or small) $\mu$. 
This would be extremely interesting, due to the opportunity to directly compare the obtained results with lattice QCD
calculations, since a finite $\mu_I$ does not lead to a sign problem,
and would give the opportunity to study $R$ in presence
of potentially two condensates, namely a pion condensate
beside the chiral condensate.
We plan to address these issues in the near future.

\begin{acknowledgements}
M. R. acknowledges 
John Petrucci for inspiration. M. R. and F. M. 
acknowledges Gabriele Parisi for numerous discussions.
F. M. acknowledges discussions with D. Rischke. This work has been partly funded by the European Union – Next Generation EU through the research grant number P2022Z4P4B “SOPHYA - Sustainable Optimised PHYsics Algorithms: fundamental physics to build an advanced society" under the program PRIN 2022 PNRR of the Italian Ministero dell’Università e Ricerca (MUR).

\end{acknowledgements}

\bibliography{biblio}

\end{document}